\documentclass[aps,prl,showpacs]{revtex4}
\usepackage{graphicx}
\usepackage{amsfonts,amssymb,amsmath}
\usepackage{bm}

\begin{document}

\title{Algebraic treatments of the problems of the spin-$1/2$ particles in
the one and two-dimensional geometry: a systematic study}
\date{\today}
 
\author{Ramazan Ko\c{c}}
\email{koc@gantep.edu.tr}
\affiliation{Department of Physics, Faculty of Engineering 
University of Gaziantep,  27310 Gaziantep, Turkey}
\author{Hayriye T\"{u}t\"{u}nc\"{u}ler}
\email{tutunculer@gantep.edu.tr}
\affiliation{Department of Physics, Faculty of Engineering 
University of Gaziantep,  27310 Gaziantep, Turkey}

\author{Mehmet Koca}
\email{kocam@squ.edu.om}
\affiliation{Department of Physics, College of Science,
Sultan Qaboos University, PO Box 36  \\
Al-Khod 123, Sultanete of Oman}
\author{Eser ol\u{g}ar}
\email{olgar@gantep.edu.tr}
\affiliation{Department of Physics, Faculty of Engineering 
University of Gaziantep,  27310 Gaziantep, Turkey}

\begin{abstract}
We consider solutions of the $2\times 2$ matrix Hamiltonians of the physical
systems within the context of the $su(2)$ and $su(1,1)$ Lie algebra. Our
technique is relatively simple when compared with the others and treats
those Hamiltonians which can be treated in a unified framework of the $%
Sp(4,R)$ algebra. The systematic study presented here reproduces a number of
earlier results in a natural way as well as leads to a novel findings.
Possible generalizations of the method are also suggested.
\end{abstract}
\maketitle

\section{Introduction}

During the last decade a great deal of attention has been paid to examine
different quantum optical models. Recently some algebraic techniques which
improve both analytical and numerical solutions of the problems, have been
suggested and developed for some quantum optical systems\cite%
{klim,kara1,kara2,kumar,wunsce,buzek,ban,chen,balan}. In general, the study
of two level-systems, in a one and two-dimensional geometry, coupled to
bosonic modes has been the subject of intense attention because of its
extensive applicability in the various fields of physics\cite%
{judd,reik,ito,mosh,koc2,tur,jaynes,rashba}.

The most general form of the Hamiltonian of a two level system in two
dimensional geometry can be expressed as%
\begin{equation}
H=H_{0}+\beta \sigma _{0}+\left( \kappa _{1}a+\kappa _{2}a^{+}+\kappa
_{3}b+\kappa _{4}b^{+}\right) \sigma _{+}+\left( \gamma _{1}a+\gamma
_{2}a^{+}+\gamma _{3}b+\gamma _{4}b^{+}\right) \sigma _{-}  \label{q1}
\end{equation}%
where $H_{0}=\hbar \omega _{1}a^{+}a+\hbar \omega _{2}b^{+}b$, $\sigma
_{0},\sigma _{+}$ and $\sigma _{-}$ are usual Pauli matrices, $a,b$ and $%
a^{+},b^{+}$ are bosonic annihilation and creation operators, respectively
and $\omega _{i},\beta ,\kappa _{i}$ and $\gamma _{1}$ are physical
constants. The Hamiltonian (\ref{q1}) includes various physical Hamiltonians
depending on the choice of the parameters. For instance, when $\omega
_{1}=\omega _{2}=\omega $, $\kappa _{2}=\kappa _{3}=\gamma _{1}=\gamma _{4}=0
$ and $\kappa _{1}=\kappa _{4}=\gamma _{2}=\gamma _{3}=\kappa ,$ the
Hamiltonian, $H,$ reduces to the $E\otimes \varepsilon $ Jahn-Teller (JT)
Hamiltonian\cite{reik,koc1}, when $\omega _{1}=\omega _{2}=\omega $, $\kappa
_{2}=\kappa _{3}=\gamma _{1}=\gamma _{4}=0$ and $\kappa _{1}=-\kappa
_{4}=\gamma _{2}=-\gamma _{3}=\kappa ,$ the Hamiltonian, $H,$ becomes the
Hamiltonians of quantum dots including spin-orbit coupling\cite{rashba,koc3}%
. One can also obtain Jaynes-Cummings (JC) Hamiltonian\cite{jaynes},
modified JC Hamiltonian\cite{koc4} as well as many other interesting
physical Hamiltonians by an appropriate choices of the parameters, $\omega
_{i},\kappa _{i},$ and $\gamma _{i}$ in (\ref{q1}). There exist a relatively
large number of different approaches for the solutions of the eigenvalue
problems in the literature; however we present here a systematic and a
unified treatment for the determination of the eigenvalues and
eigenfunctions of  (\ref{q1}), in the context of the $su(2)$ and $su(1,1)$
Lie algebra. Furthermore, we develop an algorithm and routines to implement
an exact solution scheme for determining eigenvalues and eigenfunctions of
the Hamiltonian $H$.

It is well-known that the Lie algebraic techniques are very powefull in
describing many physical problems while improving both analytical and
numerical solutions as well as understanding the nature of \ physical
structures. In this paper we concentrate our attention to the solution of
the Hamiltonian, $H,$ by constructing the proper realization of the algebras 
$su(2)$ and $su(1,1)$. In a straightforward way, furthermore, we shall see
that, the Hamiltonian (\ref{q1}) automatically leads to $su(2)$ or $su(1,1)$
algebras depending on the choices of the parameters $\kappa _{i},$ and $%
\gamma _{i}$. We also note that the algebras $su(2)$ and $su(1,1)$
describing symmetry of the Hamiltonian, $H$, can be imbedded into a larger
algebra that contains both\cite{gursey,vyb}. This algebra is $Sp(4,R)$ that
provides a unified treatment of the various approaches to the solution of
the such problems.

The paper is organized as follows. In section II we discuss the bosonisation
of the physical Hamiltonians whose original forms are given as differential
operators. In section III we briefly review the properties of $su(2)$ and $%
su(1,1)$ Lie algebras and introduce their bosonic realizations which we need
to solve the various Hamiltonians. Our main procedure are presented in
section IV, where we deal with the solution of the Hamiltonian (\ref{q1}).
The results are contained in section V. Finally we conclude our results in
section VI.

\section{Realization of the Physical Problems by bosons}

One way to relate a Hamiltonian with an appropriate Lie algebra is to
construct its bosonic and fermionic representation. We are interested in the
two-level system in a one and two-dimensional geometry, whose Hamiltonians
are given in terms of bosons-fermions or matrix-differential equations.
Therefore, it is worth to express a suitable differential realizations of
the bosons. By the use of the differential realization of the operators  one
can easily find the connection between boson-fermion and matrix differential
equation formalism of the Hamiltonians. To this end, let us start by
introducing the following differential realizations of the boson operators:%
\begin{eqnarray}
a^{+} &=&\frac{\ell }{2}(x+iy)-\frac{1}{2\ell }(\partial _{x}+i\partial
_{y}),  \notag \\
a &=&\frac{\ell }{2}(x-iy)+\frac{1}{2\ell }(\partial _{x}-i\partial _{y}), 
\notag \\
b^{+} &=&\frac{\ell }{2}(x-iy)-\frac{1}{2\ell }(\partial _{x}-i\partial
_{y}),  \label{q2} \\
b &=&\frac{\ell }{2}(x+iy)+\frac{1}{2\ell }(\partial _{x}+i\partial _{y}) 
\notag
\end{eqnarray}%
where $\ell =\sqrt{\frac{m\omega }{\hbar }}$ is the length parameter and the
boson operators obey the usual commutation relations%
\begin{equation}
\left[ a,a^{+}\right] =\left[ b,b^{+}\right] =1;\quad \left[ a,b^{+}\right] =%
\left[ b,a^{+}\right] =\left[ a,b\right] =\left[ a^{+},b^{+}\right] =0.
\label{q3}
\end{equation}%
In principle, if a Hamiltonian is expressed by boson operators, one could
rely directly on the known formulae of the action of boson operators on a
state with a defined number of particles without solving differential
equations. Apart from the mentioned method, sometimes the Hamiltonians can
be put in a simple form by using the transformation properties of the
bosons. Now, we briefly discuss the bosonic construction of the various
Hamiltonians.

\subsection{Hamiltonians of quantum dots including spin-orbit coupling}

The origin of the Rashba spin-orbit coupling in quantum dots due to the lack
of inversion symmetry which causes a local electric field perpendicular to
the plane of heterostructure. In literature the Hamiltonian has been
formalized in the coordinate-momentum space, leading to a matrix
differential equation. The Hamiltonian representing the Rashba spin orbit
coupling for an electron in a quantum dot can be expressed as\cite{rashba}%
\begin{equation}
H_{R}=\frac{\lambda _{R}}{\hbar }\left( p_{y}\sigma _{x}-p_{x}\sigma
_{y}\right)   \label{q4}
\end{equation}%
where $\lambda _{R}$ represents the strength of the spin orbit coupling,
which can be adjusted by changing the asymmetry of the quantum well via
external electric field. Here the matrices $\sigma _{x},$ and $\sigma _{y}$
are Pauli matrices. We assume that the electron is confined in a parabolic
potential%
\begin{equation}
V=\frac{1}{2}m^{\ast }\omega _{0}^{2}(x^{2}+y^{2})  \label{q5}
\end{equation}%
here $m^{\ast }$ is the effective mass of the electron and $\omega _{0}$ is
the confining potential frequency. The Hamiltonian describing an electron in
two-dimensional quantum dot takes the form%
\begin{equation}
H=\frac{1}{2m^{\ast }}\left( P_{x}^{2}+P_{y}^{2}\right) +\frac{1}{2}g\mu
B\sigma _{0}+V+H_{R}.  \label{q6}
\end{equation}%
The term $\frac{1}{2}g\mu B\sigma _{z}$ introduces the Zeeman splitting
between the $(+)x-$polarized spin up and $(-)x-$polarized spin down. The
factors $g$ is gyromagnetic ratio $\mu $ is the Bohr magneton. The canonical
momentum $\mathbf{P=p+eA}$ is expressed in terms of the mechanical momentum $%
\mathbf{p}=-i\hbar (\partial _{x},\partial _{y},0)$ and the vector potential 
$\mathbf{A}$ can be related to\ the magnetic field $\mathbf{B=\nabla }\times 
\mathbf{A}$. The choice of symmetric gauge vector potential $\mathbf{A}%
=B/2(-y,x,0)$, leads to the following Hamiltonian%
\begin{eqnarray}
H_{dot} &=&-\frac{\hbar ^{2}}{2m^{\ast }}\left( \frac{\partial ^{2}}{%
\partial x^{2}}+\frac{\partial ^{2}}{\partial y^{2}}\right) +\frac{1}{2}%
m^{\ast }\omega ^{2}(x^{2}+y^{2})+  \notag \\
&&\frac{1}{2}i\hbar \omega _{c}\left( y\frac{\partial }{\partial x}-x\frac{%
\partial }{\partial y}\right) +\frac{1}{2}g\mu B\sigma _{0}+H_{R}  \label{q7}
\end{eqnarray}%
where $\omega _{c}=eB/m^{\ast }$, stands for the cyclotron frequency of the
electron, $\omega =\sqrt{\omega _{0}^{2}+\left( \frac{\omega _{c}}{2}\right)
^{2}}$ is the effective frequency. The Hamiltonian $H_{dot}$ describing a
two-level fermionic subsystem coupled to two boson modes can be expressed as:%
\begin{eqnarray}
H &=&\hbar \omega (a^{+}a+b^{+}b+1)+\frac{\hbar \omega _{c}}{2}\left(
a^{+}a-b^{+}b\right) -  \notag \\
&&\sqrt{\frac{m\omega }{4\hbar }}\lambda _{R}\left[ (b^{+}-a)\sigma
_{+}+(b-a^{+})\sigma _{-}\right] +\frac{1}{2}g\mu B\sigma _{0}  \label{q8}
\end{eqnarray}%
The Pauli matrices are given by%
\begin{equation}
\sigma _{+}=\left( 
\begin{array}{cc}
0 & 1 \\ 
0 & 0%
\end{array}%
\right) ;\quad \sigma _{-}=\left( 
\begin{array}{cc}
0 & 0 \\ 
1 & 0%
\end{array}%
\right) ;\quad \sigma _{0}=\left( 
\begin{array}{cc}
-1 & 0 \\ 
0 & 1%
\end{array}%
\right)   \label{q9}
\end{equation}%
It is worth to point out here that the success of our construction leads to
the connection between $H_{dot}$ and $H.$

\subsection{$E\times \protect\varepsilon $ Jahn-Teller Hamiltonian}

The $E\times \varepsilon $ JT Hamiltonian describing a two level fermionic
subsystem coupled to two boson modes can be expressed\cite{reik} as%
\begin{equation}
H_{JT}=-\frac{\hbar ^{2}}{2m}\left( \frac{\partial ^{2}}{\partial x^{2}}+%
\frac{\partial ^{2}}{\partial y^{2}}\right) +\frac{1}{2}m^{\ast }\omega
_{0}^{2}(x^{2}+y^{2})+\frac{\mu }{2}\sigma _{0}+\kappa \left[ (x+iy)\sigma
_{+}+(x-iy)\sigma _{-}\right] .  \label{q10}
\end{equation}%
where $\mu $ is the level separation and $\kappa $ is the coupling
constants. In terms of bosonic operators we can easily obtain%
\begin{equation}
H_{JT}=\hbar \omega (a^{+}a+b^{+}b)+\frac{\mu }{2}\sigma _{0}+\sqrt{\frac{%
m\omega }{4\hbar }}\kappa \left[ (a+b^{+})\sigma _{+}+(a^{+}+b)\sigma _{-}%
\right] .  \label{q11}
\end{equation}%
The JT problem is an old one and the complete description of the isolated
exact solution can be found in literature. Recently it has been proven that
the JT problem possesses $osp(2,2)$ symmetry and it is one of the recently
discovered quasi-exactly solvable problem \cite{koc1}.

\subsection{Dirac Oscillator}

The $(2+1)$ dimensional Dirac equation for free particle of mass $m$ in
terms of two component spinors $\psi ,$ can be written as\cite{ito,mosh}%
\begin{equation}
E\psi =\left( \sum_{i=1}^{2}c\sigma _{i}p_{i}+mc^{2}\sigma _{0}\right) \psi .
\label{q11a}
\end{equation}%
The momentum operator $p_{i},$ is differential operator $\mathbf{p}=-i\hbar
(\partial _{x},\partial _{y})$ and the 2D Dirac oscillator can be
constructed by changing the momentum $\mathbf{p}\rightarrow \mathbf{p}%
-im\omega \sigma _{0}\mathbf{r}$. Then the Dirac equation (\ref{q11a}) takes
the form%
\begin{equation}
\left( E-mc^{2}\sigma _{0}\right) \psi =c\left[ (p_{x}-ip_{y})-im\omega
(x-iy)\right] \sigma _{+}+c\left[ (p_{x}+ip_{y})-im\omega (x+iy)\right]
\sigma _{-}.  \label{q11b}
\end{equation}%
After some straightforward treatment we obtain the bosonic form of the Dirac
oscillator:%
\begin{equation}
\left( E-mc^{2}\sigma _{0}\right) \psi =2ic\sqrt{m\omega \hbar }\left[
a\sigma _{+}+a^{+}\sigma _{-}\right] \psi .  \label{q11c}
\end{equation}%
An immediate practical consequence of these results is that the Lie
algebraic structure of the Hamiltonians can easily be determined.

In addition to those Hamiltonians which we have already bosonised, there
exist the Hamiltonians given in terms of bosons and fermions, namely, JC
Hamiltonians which can also be treated in the same manner presented in this
paper. One of them is known as JC Hamiltonian without rotating wave
approximation is given by%
\begin{equation}
H_{JC}=\hbar \omega a^{+}a+\frac{\hbar \omega _{0}}{2}\sigma _{0}+\kappa
\left( \sigma _{+}+\sigma _{-}\right) \left( a^{+}+a\right) .  \label{q11d}
\end{equation}%
The other is known as JC Hamiltonian with rotating wave approximation (RWA)
can be expressed as%
\begin{equation}
H_{JC}^{RWA}=\hbar \omega a^{+}a+\frac{\hbar \omega _{0}}{2}\sigma
_{0}+\kappa \left( \sigma _{+}a+\sigma _{-}a^{+}\right)   \label{q11e}
\end{equation}%
which can exactly be solved. When single two-level atom is placed in the
common domain of two cavities interacting with two quantized modes, the
Hamiltonian of a such system can be obtained from the modification of the JC
Hamiltonian and it is given by%
\begin{equation}
H_{MJC}=\hbar \omega a^{+}a+b^{+}b+\hbar \omega _{0}\sigma _{0}+\left(
\lambda _{1}a+\lambda _{2}b\right) \sigma _{+}+\left( \lambda
_{1}a^{+}+\lambda _{2}b^{+}\right) \sigma _{-}.  \label{q11f}
\end{equation}%
In a similar manner the two dimensional Hamiltonians can be bosonised and as
it will be shown that their algebraic structure can be easily determined.
Now, we briefly review construction of the bosonic representations of the
Lie algebras $su(2)$ and $su(1,1)$.

\section{Realizations of $su(2)$ and $su(1,1)$ by boson operators}

It is well known that if the Hamiltonians characterized by single or double
boson operators, then the simplest way to find the symmetry algebra of the
corresponding Hamiltonian is that to construct the single or double boson
realizations of the algebras. In this section we introduce some basic boson
realizations of $su(1,1)$ and $su(2)$ which we need to solve the
Hamiltonians given in previous section.

\subsection{Realizations of $su(2)$}

The $su(2)$ algebra can be constructed by two mode bosons in two dimensions
by introducing the following three generators%
\begin{equation}
J_{+}=a^{+}b;\quad J_{-}=b^{+}a;\quad J_{0}=\frac{1}{2}\left(
a^{+}a-b^{+}b\right)  \label{q12}
\end{equation}%
satisfy the commutation relations%
\begin{equation}
\left[ J_{0},J_{\pm }\right] =\pm J_{\pm };\quad \left[ J_{+},J_{-}\right]
=2J_{0}.  \label{q13a}
\end{equation}%
The number operator which commutes with the generators of the $su(2)$
algebra is given by 
\begin{equation}
N=a^{+}a+b^{+}b.  \label{q14}
\end{equation}%
Casimir invariant of $su(2)$ can be related to $N$ by%
\begin{equation}
C=\frac{1}{4}N(N+2).  \label{q15a}
\end{equation}%
The eigenvalues of $C$ are given by%
\begin{equation}
\left\langle C\right\rangle =j(j+1).  \label{q16}
\end{equation}%
It is obvious that the irreducible representations of $su(2)$ can be
characterized by the total boson number $N=2j$. The application of the
realization (\ref{q12}) on a set of \ $2j+1$ states, leads to the $(2j+1)$%
-dimensional unitary irreducible representation for each$j=0,1/2,1,\cdots $.
If the basis states $\left| j,m\right\rangle $ $(m=j,j-1,....-j)$, then the
action of the operators on the basis is given by%
\begin{eqnarray}
J_{0}\left| j,m\right\rangle &=&m\left| j,m\right\rangle  \notag \\
J_{\pm }\left| j,m\right\rangle &=&\sqrt{(j\mp m)(j\pm m+1)}\left| j,m\pm
1\right\rangle  \label{q16a} \\
C\left| j,m\right\rangle &=&j(j+1)\left| j,m\right\rangle  \notag
\end{eqnarray}

Furthermore, the single boson realizations of $su(2)$ algebra can be
constructed by defining three operators%
\begin{equation}
J_{+}^{\prime }=\sqrt{2j-N^{\prime }}a;\quad J_{-}^{\prime }=a^{+}\sqrt{%
2j-N^{\prime }};\quad J_{0}^{\prime }=j-N^{\prime };\quad N^{\prime }=a^{+}a
\label{q17}
\end{equation}%
also satisfy the commutation relations of the $su(2)$ algebra.

\subsection{Realizations of $su(1,1)$ by boson operators}

The Lie algebra $su(1,1)$ possesses interesting realizations by bosons and
more appropriate to solve the many physical problems. Using the set of boson
operators (\ref{q2}) we introduce the three operators%
\begin{equation}
K_{+}=a^{+}b^{+};\quad K_{-}=ab;\quad K_{0}=\frac{1}{2}\left(
a^{+}a+b^{+}b+1\right) .  \label{q18}
\end{equation}%
The reader can easily check that the $K^{\prime }s$ satisfy the $su(1,1)$
commutation relations%
\begin{equation}
\left[ K_{0},K_{\pm }\right] =\pm K_{\pm };\quad \left[ K_{+},K_{-}\right]
=-2K_{0}.  \label{q19}
\end{equation}%
While in the $su(2)$ case the number operator was the sum, $N$, of the boson
number operators, in the present case it is the operator%
\begin{equation}
M=a^{+}a-b^{+}b,  \label{q20}
\end{equation}%
which is the difference of the number operators. The casimir invariant of
the $su(1,1)$ is related to $M$ by%
\begin{equation}
C=\frac{1}{4}(1+M)(1-M).  \label{q21}
\end{equation}%
Therefore if the eigenvalues of the operator $C$ is $k(1-k)$ we find that $%
M=(1-2k).$ Consequently, the action of the realization (\ref{q18}) on the
states $\left| k,n\right\rangle ,(n=0,1,2,\cdots ),$ leads to infinite
dimensional unitary irreducible representation, so called positive
representation $D^{+}(k)$, corresponds to any $k=1/2,1,3/2,\cdots .$ 
\begin{eqnarray}
K_{0}\left| k,n\right\rangle  &=&(k+n)\left| k,n\right\rangle   \notag \\
K_{+}\left| k,n\right\rangle  &=&\sqrt{(2k+n)(n+1)}\left| k,n+1\right\rangle 
\notag \\
K_{-}\left| k,n\right\rangle  &=&\sqrt{(2k+n-1)n}\left| k,n-1\right\rangle 
\label{q21a} \\
C\left| k,n\right\rangle  &=&k(1-k)\left| k,n\right\rangle   \notag
\end{eqnarray}

We will end this section by introducing two different single mode boson
realizations of the $su(1,1)$ algebra. One of them can be constructed by the
operators:%
\begin{equation}
L_{+}=\frac{1}{2}a^{+2};\quad L_{-}=\frac{1}{2}a^{2};\quad L_{0}=\frac{1}{2}%
\left( a^{+}a+\frac{1}{2}\right) ;\quad M^{\prime }=a^{+}a;  \label{q22}
\end{equation}%
For the single-mode bosonic realization of $su(1,1)$ that we require here,
the Bargmann index $k$ is equal to either $1/4$ or $3/4$ which split the
Hilbert space of the boson space into two independent subspace. The other
realization is%
\begin{equation}
S_{+}=a^{+}\sqrt{M^{\prime }+2k};\quad S_{-}=\sqrt{M^{\prime }+2k}a;\quad
S_{0}=M^{\prime }+k;\quad k>0.  \label{q23}
\end{equation}%
It will be shown that in particular the realizations (\ref{q18}) and (\ref%
{q22}) play dominant role on the solution of the Hamiltonian (\ref{q1}).

\section{Method}

In our method the bosonised Hamiltonians are connected with $su(1,1)$ and $%
su(2)$ , as well as $Sp(4,R)$ Lie algebras. This connection opens the way to
an algebraic treatment of a large class of physical Hamiltonians of
practical interest. In this section we present a general procedure to solve
the Hamiltonian (\ref{q1}). Consider the eigenvalue equation 
\begin{equation}
H\psi =E\psi   \label{q23a}
\end{equation}%
where $\psi $ is two component wavefunction and $E$ is eigenvalues of the
Hamiltonian $H$. Consequently the Hamiltonian (\ref{q1}) can be written as 
\begin{subequations}
\begin{eqnarray}
\left( H_{0}-E-\beta \right) \psi _{1}+\left( \kappa _{1}a+\kappa
_{2}a^{+}+\kappa _{3}b+\kappa _{4}b^{+}\right) \psi _{2} &=&0  \label{q24a}
\\
\left( H_{0}-E+\beta \right) \psi _{2}+\left( \gamma _{1}a+\gamma
_{2}a^{+}+\gamma _{3}b+\gamma _{4}b^{+}\right) \psi _{1} &=&0.  \label{q24b}
\end{eqnarray}%
These coupled equations may be solved by using various techniques. In here
we follow a new strategies. In the first step we eliminate $\psi _{2}$ (or $%
\psi _{1}$) between the above equation 
\end{subequations}
\begin{equation}
\psi _{2}=-\left( H_{0}-E+\beta \right) ^{-1}\left( \gamma _{1}a+\gamma
_{2}a^{+}+\gamma _{3}b+\gamma _{4}b^{+}\right) \psi _{1}.  \label{q25}
\end{equation}%
Substituting (\ref{q25}) in to (\ref{q24a}) we obtain the following equation%
\begin{eqnarray}
&&\left( H_{0}-E-\beta \right) \psi _{1}-  \notag \\
&&\left( \kappa _{1}a+\kappa _{2}a^{+}+\kappa _{3}b+\kappa _{4}b^{+}\right)
\left( H_{0}-E+\beta \right) ^{-1}\left( \gamma _{1}a+\gamma
_{2}a^{+}+\gamma _{3}b+\gamma _{4}b^{+}\right) \psi _{1}=0.  \label{q26}
\end{eqnarray}%
The last equation can be solved by performing a suitable realizations of the 
$su(1,1)$ and $su(2)$ algebra. In our formalism, in order to obtain an
adequate form of the (\ref{q26}), in the next step, we use the relations%
\begin{eqnarray}
a\left( H_{0}-E+\beta \right) ^{-1} &=&\left( H_{0}-E+\beta -\omega
_{1}\right) ^{-1}a  \notag \\
a^{+}\left( H_{0}-E+\beta \right) ^{-1} &=&\left( H_{0}-E+\beta +\omega
_{1}\right) ^{-1}a^{+}  \notag \\
b\left( H_{0}-E+\beta \right) ^{-1} &=&\left( H_{0}-E+\beta -\omega
_{2}\right) ^{-1}b  \label{q27} \\
b^{+}\left( H_{0}-E+\beta \right) ^{-1} &=&\left( H_{0}-E+\beta +\omega
_{2}\right) ^{-1}b^{+}  \notag
\end{eqnarray}%
by setting $\omega _{1}=\omega _{2}=\omega $, we obtain the following
general expression%
\begin{eqnarray}
&&\left( H_{0}-E+\beta +\hbar \omega \right) \left( H_{0}-E+\beta -\hbar
\omega \right) \left( H_{0}-E-\beta \right) \psi _{1}=  \notag \\
&&\kappa _{1}\left( H_{0}-E+\beta +\hbar \omega \right) \left( \gamma
_{1}a^{2}+\gamma _{2}aa^{+}+\gamma _{3}ab+\gamma _{4}ab^{+}\right) \psi _{1}+
\notag \\
&&\kappa _{2}\left( H_{0}-E+\beta -\hbar \omega \right) \left( \gamma
_{1}a^{+}a+\gamma _{2}a^{+2}+\gamma _{3}a^{+}b+\gamma _{4}a^{+}b^{+}\right)
\psi _{1}+  \notag \\
&&\kappa _{3}\left( H_{0}-E+\beta +\hbar \omega \right) \left( \gamma
_{1}ba+\gamma _{2}ba^{+}+\gamma _{3}b^{2}+\gamma _{4}bb^{+}\right) \psi _{1}+
\notag \\
&&\kappa _{4}\left( H_{0}-E+\beta -\hbar \omega \right) \left( \gamma
_{1}b^{+}a+\gamma _{2}b^{+}a^{+}+\gamma _{3}b^{+}b+\gamma _{4}b^{+2}\right)
\psi _{1}.  \label{q28}
\end{eqnarray}%
It can be checked easily that for some certain values of $\kappa _{i}$, $%
\gamma _{i}$ and $\beta $ the equation can be solved in the framework of $%
su(1,1)$ or $su(2)$ Lie algebra techniques and the Hamiltonian can be
related to the various physical Hamiltonians of interest. The resulting
Hamiltonians are summarized in the \ref{Table}. 
\begin{table}[t]
\begin{tabular}{|c|c|c|c|c|c|c|c|c|c|c|c|c|c|}
\hline
\multicolumn{1}{|c|}{Related} & \multicolumn{11}{|c}{Parameters} & 
\multicolumn{2}{|c|}{Symmetry} \\ \hline
\multicolumn{1}{|c|}{Hamiltonian} & $\kappa _{1}$ & $\kappa _{2}$ & $\kappa
_{3}$ & $\kappa _{4}$ & $\gamma _{1}$ & $\gamma _{2}$ & \multicolumn{1}{|c|}{%
$\gamma _{3}$} & $\gamma _{4}$ & $\omega _{1}$ & $\omega _{2}$ & $\beta $ & 
\multicolumn{2}{|c|}{Group} \\ \hline
$H_{JT}$ & $\kappa ^{\prime }$ & $0$ & $0$ & $\kappa ^{\prime }$ & $0$ & $%
\kappa ^{\prime }$ & \multicolumn{1}{|c|}{$\kappa ^{\prime }$} & $0$ & $%
\omega $ & $\omega $ & $\frac{\mu }{2}$ & $su(1,1)$ & $Eq.(\ref{q18})$ \\ 
\hline
$H_{dot}$ & $\lambda ^{\prime }$ & $0$ & $0$ & $-\lambda ^{\prime }$ & $0$ & 
$\lambda ^{\prime }$ & \multicolumn{1}{|c|}{$-\lambda ^{\prime }$} & $0$ & $%
\omega $ & $\omega $ & $\frac{\mu }{2}gB$ & $su(1,1)$ & $Eq.(\ref{q18})$ \\ 
\hline
$H_{JC}$ & $\kappa $ & $\kappa $ & $0$ & $0$ & $\kappa $ & $\kappa $ & 
\multicolumn{1}{|c|}{$0$} & $0$ & $\omega $ & $0$ & $\frac{\hbar \omega _{0}%
}{2}$ & $su(1,1)$ & $Eq.(\ref{q20})$ \\ \hline
$H_{JC}^{RWA}$ & $\kappa $ & $0$ & $0$ & $0$ & $0$ & $\kappa $ & 
\multicolumn{1}{|c|}{$0$} & $0$ & $\omega $ & $0$ & $\frac{\hbar \omega _{0}%
}{2}$ & $su(1,1)$ & $Eq.(\ref{q20})$ \\ \hline
$H_{MJC}$ & $\lambda _{1}$ & $0$ & $\lambda _{2}$ & $0$ & $0$ & $\lambda
_{1} $ & \multicolumn{1}{|c|}{$0$} & $\lambda _{2}$ & $\omega $ & $\omega $
& $\hbar \omega _{0}$ & $su(2)$ & $Eq.(\ref{q12})$ \\ \hline
$H_{Dirac}$ & $\kappa ^{\prime \prime }$ & $0$ & $0$ & $0$ & $0$ & $\kappa
^{\prime \prime }$ & \multicolumn{1}{|c|}{$0$} & $0$ & $0$ & $0$ & $mc^{2}$
& $su(1,1)$ & $Eq.(\ref{q20})$ \\ \hline
\end{tabular}%
\caption{List of the Hamiltonians depending on the choices of the parameters
of Hamiltonian (\ref{q1}). The corresponding Hamiltonians are illustrated in
the first column and their algebras and appropriate realizations are
illustrated in the last column. The parameters $\protect\kappa ^{\prime }=%
\protect\sqrt{\frac{m\protect\omega }{4\hbar }}\protect\kappa ,\quad \protect%
\kappa ^{\prime \prime }=2ic\protect\sqrt{m\protect\omega \hbar },\quad 
\protect\lambda ^{\prime }=\protect\sqrt{\frac{m\protect\omega }{4\hbar }}%
\protect\lambda _{R}.$}
\label{tab:b}
\end{table}

\section{Results and Discussions}

The results of our investigation are that, for the Hamiltonians whose list
given in Table 1, are demonstrated in two categories; exactly \cite%
{gursey,levai} and quasi-exactly solvable Hamiltonians\cite%
{turb1,bender,brih,shif}. It will be shown that, the rotating wave
approximated JC Hamiltonian, Dirac oscillator and modified JC Hamiltonian,
as they expected, exactly solvable, while JT, JC and quantum dot
Hamiltonians are quasi-exactly solvable.

\subsection{Exactly solvable Hamiltonians}

The JC Hamiltonian with the rotating wave approximation can be obtained by
setting $\kappa _{2}=$ $\kappa _{3}=\kappa _{4}=\gamma _{1}=\gamma
_{3}=\gamma _{4}=0,\kappa _{1}=\gamma _{2}=\kappa $ and $\beta =\frac{\hbar
\omega _{0}}{2}.$ In this case the formula (\ref{q28}) takes the form: 
\begin{eqnarray}
&&\left( H_{0}-E+\frac{\hbar \omega _{0}}{2}+\hbar \omega \right) \left(
H_{0}-E+\frac{\hbar \omega _{0}}{2}-\hbar \omega \right) \left( H_{0}-E-%
\frac{\hbar \omega _{0}}{2}\right) \psi _{1}=  \notag \\
&&\kappa ^{2}\left( H_{0}-E+\beta +\hbar \omega \right) aa^{+}\psi _{1}.
\label{q29}
\end{eqnarray}%
The algebraic form of the Hamiltonian can be obtained by combining (\ref%
{q29a}) and the $su(1,1)$ realization given in (\ref{q22}), yields%
\begin{equation}
\left( \hbar \omega M^{\prime }-E+\frac{\hbar \omega _{0}}{2}-\hbar \omega
\right) \left( \hbar \omega M^{\prime }-E-\frac{\hbar \omega _{0}}{2}\right)
\psi _{1}=\kappa ^{2}\left( M^{\prime }+1\right) \psi _{1}  \label{q30a}
\end{equation}%
If the wavefunction $\psi _{1}=\left| \frac{1}{4},n\right\rangle $ then the
action of the (\ref{q30a}) on the state leads to the following expression
for the eigenvalues of the JC Hamiltonian%
\begin{equation}
E=\left( n-\frac{1}{2}\right) \hbar \omega -\frac{\hbar \omega _{0}}{2}\pm 
\sqrt{\hbar ^{2}\omega ^{2}+4\kappa ^{2}(n+1)}  \label{q31a}
\end{equation}%
It is obvious that when the coupling constant $\kappa $ is zero then the
result is the eigenvalues of the simple harmonic oscillator.

The other exactly solvable problem is the Dirac oscillator. It can be
obtained by setting the parameters $\omega _{1}=\omega _{2}=\kappa _{2}=$ $%
\kappa _{3}=\kappa _{4}=\gamma _{1}=\gamma _{3}=\gamma _{4}=0,\quad \kappa
_{1}=\gamma _{2}=2ic\sqrt{m\omega \hbar }$ and $\beta =mc^{2}$, yields the
following expression:%
\begin{equation}
\left( -E+mc^{2}\right) \left( -E+mc^{2}\right) \left( -E-mc^{2}\right) \psi
_{1}=-4mc^{2}\hbar \omega \left( -E+mc^{2}\right) aa^{+}\psi _{1}
\label{q32a}
\end{equation}%
The algebraic form of the equation, with the realization (\ref{q22}), is
given by 
\begin{equation}
\left( -E+mc^{2}\right) \left( -E-mc^{2}\right) \psi _{1}=-4mc^{2}\hbar
\omega (M^{\prime }+1)\psi _{1}  \label{q33}
\end{equation}%
Then we obtain the following energy values for the Dirac oscillator:%
\begin{equation}
E=\pm \sqrt{m^{2}c^{4}-4\hbar \omega mc^{2}(n+1)}  \label{q34}
\end{equation}%
which is an exact eigenvalues of the Dirac oscillator.

The last exactly solvable problem we consider here is the MJC model. The
Hamiltonian can be obtained by choosing the parameters: $\kappa _{2}=\kappa
_{4}=\gamma _{1}=\gamma _{3}=0$, $\kappa _{1}=\gamma _{2}=\lambda _{1},$ $%
\kappa _{3}=\gamma _{4}=\lambda _{2},$ and $\beta =\hbar \omega _{0}$ then (%
\ref{q28}) takes the form:%
\begin{eqnarray}
&&\left( H_{0}-E+\hbar \omega _{0}+\hbar \omega \right) \left( H_{0}-E+\hbar
\omega _{0}-\hbar \omega \right) \left( H_{0}-E-\hbar \omega _{0}\right)
\psi _{1}=  \notag \\
&&\lambda _{1}\left( H_{0}-E+\hbar \omega _{0}+\hbar \omega \right) \left(
\lambda _{1}aa^{+}+\lambda _{2}ab^{+}\right) \psi _{1}+  \label{q35} \\
&&\lambda _{2}\left( H_{0}-E+\hbar \omega _{0}+\hbar \omega \right) \left(
\lambda _{1}ba^{+}+\lambda _{2}bb^{+}\right) \psi _{1}  \notag
\end{eqnarray}%
We insert (\ref{q12}) and (\ref{q14}), the realization of $su(2)$, into (\ref%
{q35}) we obtain the following expressions

\begin{eqnarray}
&&\left( \hbar \omega N-E+\hbar \omega _{0}-\hbar \omega \right) \left(
\hbar \omega N-E-\hbar \omega _{0}\right) \psi _{1}=  \notag \\
&&\left( \left( \lambda _{1}^{2}+\lambda _{2}^{2}\right) \left( 1+\frac{N}{2}%
\right) +(\lambda _{1}^{2}-\lambda _{2}^{2})J_{0}+\lambda _{1}\lambda
_{2}(J_{+}+J_{-})\right) \psi _{1}  \label{q36a}
\end{eqnarray}%
The above equation is not diagonal, but it can be diagonalized by similarity
transformation, by the operator%
\begin{equation}
O=e^{\frac{\alpha }{2}\left( J_{+}-J_{-}\right) }.  \label{q37a}
\end{equation}%
The action of the operator on the generators of $su(2)$ is given by%
\begin{eqnarray}
O(J_{+}+J_{-})O^{\dagger } &=&(J_{+}+J_{-})\cos \alpha +2J_{0}\sin \alpha  
\notag \\
OJ_{0}O^{\dagger } &=&J_{0}\cos \alpha -\frac{J_{+}+J_{-}}{2}\sin \alpha 
\label{q38} \\
ONO^{\dagger } &=&N.  \notag
\end{eqnarray}%
The transformations of the generators of $su(2)$ by the operator $O$ are
given by 
\begin{eqnarray}
&&\left( \hbar \omega N-E+\hbar \omega _{0}-\hbar \omega \right) \left(
\hbar \omega N-E-\hbar \omega _{0}\right) \psi _{1}=  \notag \\
&&\left( \lambda _{1}^{2}+\lambda _{2}^{2}\right) \left( 1+\frac{N}{2}%
\right) \psi _{1}+\left( (\lambda _{1}^{2}-\lambda _{2}^{2})\cos \alpha
+2\lambda _{1}\lambda _{2}\sin \alpha \right) J_{0}\psi _{1}+  \label{q39} \\
&&\left( \lambda _{1}\lambda _{2}\cos \alpha -\frac{(\lambda
_{1}^{2}-\lambda _{2}^{2})}{2}\sin \alpha \right) (J_{+}+J_{-})\psi _{1} 
\notag
\end{eqnarray}%
The Hamiltonian takes the diagonal form when the following condition hold:%
\begin{equation}
\alpha =\cos ^{-1}\left( \frac{(\lambda _{1}^{2}-\lambda _{2}^{2})}{(\lambda
_{1}^{2}+\lambda _{2}^{2})}\right) .  \label{q40}
\end{equation}%
The final form of the diagonal Hamiltonian under the condition (\ref{q40})
is given by%
\begin{eqnarray}
&&\left( \hbar \omega N-E+\hbar \omega _{0}-\hbar \omega \right) \left(
\hbar \omega N-E-\hbar \omega _{0}\right) \psi _{1}=  \notag \\
&&\left( \lambda _{1}^{2}+\lambda _{2}^{2}\right) \left( 1+\frac{N}{2}%
\right) \psi _{1}+(\lambda _{1}^{2}+\lambda _{2}^{2})J_{0}\psi _{1}
\label{q41}
\end{eqnarray}%
It is obvious that when $\psi _{1}=\left| j,m\right\rangle ,$ which is which
is the eigenstate of the operators $J_{0}$, and $N$, we obtain the following
expression for the eigenvalues of the MJC Hamiltonian:%
\begin{equation}
E=(2j-\frac{1}{2})\hbar \omega -\hbar \omega _{0}\pm \frac{1}{2}\sqrt{\hbar
^{2}\omega ^{2}+4\left( \lambda _{1}^{2}+\lambda _{2}^{2}\right) \left(
j+m+1\right) }.  \label{q42}
\end{equation}%
Without further details, we have solved various physical Hamiltonians in the
framework of the method given in section V. \ Our task is now to show the
method given in the previous section can also be applied to obtain QES of
the some physical Hamiltonians.

\subsection{QES Hamiltonians}

In recent years there has been great deal of interest in quantum optical
models which reveal new physical phenomena described by Hamiltonians
expressed in terms of the boson and fermion operators. It will be shown that
such systems can be analyzed using the method presented in this paper.
Consequently, their finite number of eigenvalues and associated
eigenfunctions can be obtained in the closed form. These systems are said to
be QES systems\cite{turb1,turb2,bender,brih,zhd,shif,alvar}.

The JC Hamiltonian without rotating wave approximation is one of the such
QES problems that can be associated to the Hamiltonian (\ref{q1}) with the
choices of the parameters: $\omega _{2}=\kappa _{3}=\kappa _{4}=\gamma
_{3}=\gamma _{4}=0,\kappa _{1}=\kappa _{2}=\gamma _{1}=\gamma _{2}=\kappa ,$ 
$\beta =\frac{\hbar \omega _{0}}{2}$ and $\omega _{1}=\omega .$ Thus the
general expression (\ref{q28}) takes the form%
\begin{eqnarray}
&&\left( H_{0}-E+\frac{\hbar \omega _{0}}{2}+\hbar \omega \right) \left(
H_{0}-E+\frac{\hbar \omega _{0}}{2}-\hbar \omega \right) \left( H_{0}-E-%
\frac{\hbar \omega _{0}}{2}\right) \psi _{1}=  \notag \\
&&\kappa ^{2}\left( H_{0}-E+\frac{\hbar \omega _{0}}{2}+\hbar \omega \right)
\left( a^{2}+aa^{+}\right) \psi _{1}+  \label{q43} \\
&&\kappa ^{2}\left( H_{0}-E+\frac{\hbar \omega _{0}}{2}-\hbar \omega \right)
\left( a^{+}a+a^{+2}\right) \psi _{1}.  \notag
\end{eqnarray}%
The JC Hamiltonian can be expressed in terms of the $su(1,1)$ generators
given in (\ref{q22}): 
\begin{eqnarray}
&&\left( \hbar \omega M^{\prime }-E+\frac{\hbar \omega _{0}}{2}+\hbar \omega
\right) \left( \hbar \omega M^{\prime }-E+\frac{\hbar \omega _{0}}{2}-\hbar
\omega \right) \left( \hbar \omega M^{\prime }-E-\frac{\hbar \omega _{0}}{2}%
\right) \psi _{1}=  \notag \\
&&\kappa ^{2}\left( \hbar \omega M^{\prime }-E+\frac{\hbar \omega _{0}}{2}%
+\hbar \omega \right) \left( 2L_{0}+M^{\prime }+1\right) \psi _{1}+
\label{q44} \\
&&\kappa ^{2}\left( \hbar \omega M^{\prime }-E+\frac{\hbar \omega _{0}}{2}%
-\hbar \omega \right) \left( 2L_{+}+M^{\prime }\right) \psi _{1}  \notag
\end{eqnarray}%
Here the Bargmann index $k$ is $\frac{1}{4}$ for the even state $\psi
_{1}=\left| \frac{1}{4},2n\right\rangle $ and $k=\frac{3}{4}$ for odd state $%
\psi _{1}=\left| \frac{3}{4},2n+1\right\rangle .$ Under the action of the
operators on the even state $\left| \frac{1}{4},2n\right\rangle \equiv
\left| 2n\right\rangle $ we obtain the following recurrence relation%
\begin{eqnarray}
&&\left( 2\hbar \omega n-E+\frac{\hbar \omega _{0}}{2}+\hbar \omega \right)
\left( 2\hbar \omega n-E+\frac{\hbar \omega _{0}}{2}-\hbar \omega \right)
\left( 2\hbar \omega n-E-\frac{\hbar \omega _{0}}{2}\right) \left|
2n\right\rangle =  \notag \\
&&\kappa ^{2}\left( 2\hbar \omega n-E+\frac{\hbar \omega _{0}}{2}+\hbar
\omega \right) \left( \sqrt{2n(2n-1)}\left| 2n-2\right\rangle +(2n+1)\left|
2n\right\rangle \right) +  \label{q45} \\
&&\kappa ^{2}\left( 2\hbar \omega n-E+\frac{\hbar \omega _{0}}{2}-\hbar
\omega \right) \left( \sqrt{(2n+1)(2n+2)}\left| 2n+2\right\rangle +2n\left|
2n\right\rangle \right)   \notag
\end{eqnarray}%
The eigenvalues can be obtained from the recurrence relation and its ground
state energy is given by%
\begin{equation}
n=0;E=\frac{1}{2}\left( -\hbar \omega \pm \sqrt{4\kappa ^{2}+\hbar
^{2}(\omega -\omega _{0})^{2}}\right) .  \label{q46}
\end{equation}%
In order to obtain the remaining eigenvalues, it is enough to perform
calculation in (\ref{q46}), but for higher values of $n$ (i.e. $n>5$) it
requires numerical treatments.

Another physically important and QES Hamiltonian is the Hamiltonian of the
quantum a dot including spin orbit coupling obtained from the formula (\ref%
{q28}), by setting the parameters: $\kappa _{2}=$ $\kappa _{3}=\gamma
_{1}=\gamma _{4}=0,$ we obtain $\kappa _{4}=-\kappa _{1}=-\gamma _{2}=\gamma
_{3}=-\sqrt{\frac{m\omega }{4\hbar }}\lambda _{R}$ and $\beta =\frac{1}{2}%
g\mu B:$%
\begin{eqnarray}
&&\left( H_{0}-E+\frac{1}{2}g\mu B+\hbar \omega \right) \left( H_{0}-E+\frac{%
1}{2}g\mu B-\hbar \omega \right) \left( H_{0}-E-\frac{1}{2}g\mu B\right)
\psi _{1}=  \notag \\
&&\frac{m\omega }{4\hbar }\lambda _{R}^{2}\left( H_{0}-E+\frac{1}{2}g\mu
B+\hbar \omega \right) \left( aa^{+}+ab\right) \psi _{1}+  \label{q47} \\
&&\frac{m\omega }{4\hbar }\lambda _{R}^{2}\left( H_{0}-E+\frac{1}{2}g\mu
B-\hbar \omega \right) \left( b^{+}a^{+}+b^{+}b\right) \psi _{1}  \notag
\end{eqnarray}%
The appropriate Lie algebra of this system is $su(1,1)$ and it can be
written in terms of the generators (\ref{q18}):

\begin{eqnarray}
&&\left( 2\hbar \omega K_{0}-E+\frac{1}{2}g\mu B\right) \left( 2\hbar \omega
K_{0}-E+\frac{1}{2}g\mu B-2\hbar \omega \right) \left( 2\hbar \omega K_{0}-E-%
\frac{1}{2}g\mu B-\hbar \omega \right) \psi _{1}=  \notag \\
&&\frac{m\omega }{4\hbar }\lambda _{R}^{2}\left( 2\hbar \omega K_{0}-E+\frac{%
1}{2}g\mu B\right) \left( K_{0}+K_{-}+\frac{M}{2}\right) \psi _{1}+
\label{q48} \\
&&\frac{m\omega }{4\hbar }\lambda _{R}^{2}\left( 2\hbar \omega K_{0}-E+\frac{%
1}{2}g\mu B-2\hbar \omega \right) \left( K_{0}+K_{+}-\frac{M+1}{2}\right)
\psi _{1}.  \notag
\end{eqnarray}%
The basis function of the (\ref{q48}) is $\psi _{1}=\left| k,n\right\rangle ,
$ by using the action of the generators on the state, we obtain the three
term recurrence relation%
\begin{eqnarray}
&&\left( 2\hbar \omega (k+n)-E_{+}\right) \left( 2\hbar \omega
(k+n)-E_{+}-2\hbar \omega \right) \left( 2\hbar \omega (k+n)-E_{-}-\hbar
\omega \right) \left| k,n\right\rangle =  \notag \\
&&\frac{m\omega }{4\hbar }\lambda _{R}^{2}\left( 2\hbar \omega
(k+n)-E_{+}\right) \left( (n+\frac{1}{2})\left| k,n\right\rangle +\sqrt{%
(2k+n-1)n}\left| k,n-1\right\rangle \right) +  \label{q49} \\
&&\frac{m\omega }{4\hbar }\lambda _{R}^{2}\left( 2\hbar \omega
(k+n)-E_{+}-2\hbar \omega \right) \left( (2k+n-1)\left| k,n\right\rangle +%
\sqrt{(2k+n)(n+1)}\left| k,n+1\right\rangle \right) .  \notag
\end{eqnarray}%
where $E_{\pm }=E\mp \frac{1}{2}g\mu B.$ The solution of the recurrence
relation for each values of $k=0,1,2,\cdots $ and $n=0,1,2,\cdots ,k$ gives
an expression for the eigenvalues of the Hamiltonian $H_{dot}.$

Our last example is the JT Hamiltonian which can also be treated as the QES
problem. In this case the parameters take the values: $\gamma _{1}=\gamma
_{4}=\kappa _{2}=\kappa _{3}=0,$ $\gamma _{2}=\gamma _{3}=\kappa _{1}=\kappa
_{4}=\sqrt{\frac{m\omega }{4\hbar }}\kappa $ and $\beta =\frac{\mu }{2}.$
Thus the eigenvalue equation takes the form: 
\begin{eqnarray}
&&\left( H_{0}-E+\frac{\mu }{2}+\hbar \omega \right) \left( H_{0}-E+\frac{%
\mu }{2}-\hbar \omega \right) \left( H_{0}-E-\frac{\mu }{2}\right) \psi _{1}=
\notag \\
&&\frac{m\omega }{4\hbar }\kappa ^{2}\left( H_{0}-E+\frac{\mu }{2}+\hbar
\omega \right) \left( aa^{+}+ab\right) \psi _{1}+  \label{q50} \\
&&\frac{m\omega }{4\hbar }\kappa ^{2}\left( H_{0}-E+\frac{\mu }{2}-\hbar
\omega \right) \left( b^{+}a^{+}+b^{+}b\right) \psi _{1}.  \notag
\end{eqnarray}%
The JT problem can also possesses $su(1,1)$ symmetry and in terms of its
generators it can be written as%
\begin{eqnarray}
&&\left( 2\hbar \omega K_{0}-E+\frac{\mu }{2}\right) \left( 2\hbar \omega
K_{0}-E+\frac{\mu }{2}-2\hbar \omega \right) \left( 2\hbar \omega K_{0}-E-%
\frac{\mu }{2}-\hbar \omega \right) \psi _{1}=  \notag \\
&&\frac{m\omega }{4\hbar }\kappa ^{2}\left( 2\hbar \omega K_{0}-E+\frac{\mu 
}{2}\right) \left( K_{-}+K_{0}+\frac{M+1}{2}\right) \psi _{1}+  \label{q51}
\\
&&\frac{m\omega }{4\hbar }\kappa ^{2}\left( 2\hbar \omega K_{0}-E+\frac{\mu 
}{2}-2\hbar \omega \right) \left( K_{+}+K_{0}-\frac{M+1}{2}\right) \psi _{1}.
\notag
\end{eqnarray}%
This algebraic equation with the basis $\psi _{1}=\left| k,n\right\rangle ,$
leads to the three term recurrence relation:%
\begin{eqnarray}
&&\left( 2\hbar \omega (k+n)-E^{+}\right) \left( 2\hbar \omega
(k+n)-E^{+}-2\hbar \omega \right) \left( 2\hbar \omega (k+n)-E^{-}-\hbar
\omega \right) \left| k,n\right\rangle =  \notag \\
&&\frac{m\omega }{4\hbar }\kappa ^{2}\left( 2\hbar \omega (k+n)-E^{+}\right)
\left( \sqrt{(2k+n-1)n}\left| k,n-1\right\rangle +(n+1)\left|
k,n\right\rangle \right) +  \label{q53} \\
&&\frac{m\omega }{4\hbar }\kappa ^{2}\left( 2\hbar \omega (k+n)-E^{+}-2\hbar
\omega \right) \left( \sqrt{(2k+n)(n+1)}\left| k,n+1\right\rangle
+(2k+n-1)\left| k,n\right\rangle \right) .  \notag
\end{eqnarray}%
where $E^{\pm }=E\mp \frac{\mu }{2}.$ The eigenvalues of the JT Hamiltonian
can be obtained by solving the recurrence relation (\ref{q53}).

As a consequence we have demonstrated that the solution of the Hamiltonian (%
\ref{q1}) can be treated within the method presented in this paper for
certain values of the parameters. Our approach is relatively simple when
compared previous approaches.

\section{Conclusion}

Here we have systematically discussed exact and QES of the Hamiltonian (\ref%
{q1}), within the framework of $su(2)$ and $su(1,1)$ Lie algebra. The
technique given here can be used in determining the spectrum of the variety
of physical systems. We have shown that our formulation leads to the exact
or quasi-exact solution of the problems of the various physical systems.

As a further work the technique can be devoloped such that one can construct 
$Sp(4,R)$ algebra which include both $su(2)$ and $su(1,1)$ algebras to study
the general Hamiltonian (\ref{q1}). The simplest way to construct this
algebra is by using boson approach discussed in section III. In addition to
the generators $J_{\pm },K_{\pm },L_{\pm },J_{0},K_{0},L_{0},N,M,$ and $%
M^{\prime },$ the operators%
\begin{equation*}
T_{+}=\frac{1}{2}b^{+2};\quad T_{-}=\frac{1}{2}b;\quad T_{0}=\frac{1}{2}%
b^{+}b+\frac{1}{4}
\end{equation*}%
forms $Sp(4,R)$ algebra. One can show that the general expression (\ref{q28}%
) can be expressed in terms of the generators of the $Sp(4,R)$ algebra: 
\begin{eqnarray*}
&&\left( \hbar \omega N-E+\beta +\hbar \omega \right) \left( \hbar \omega
N-E+\beta -\hbar \omega \right) \left( \hbar \omega N-E-\beta \right) \psi
_{1}= \\
&&\kappa _{1}\left( \hbar \omega N-E+\beta +\hbar \omega \right) \left(
\gamma _{1}L_{-}+\gamma _{2}(1+M)+\gamma _{3}K_{-}+\gamma _{4}J_{-}\right)
\psi _{1}+ \\
&&\kappa _{2}\left( \hbar \omega N-E+\beta -\hbar \omega \right) \left(
\gamma _{1}M+\gamma _{2}L_{+}+\gamma _{3}J_{+}+\gamma _{4}K_{+}\right) \psi
_{1}+ \\
&&\kappa _{3}\left( \hbar \omega N-E+\beta +\hbar \omega \right) \left(
\gamma _{1}K_{-}+\gamma _{2}J_{+}+\gamma _{3}T_{-}+\gamma _{4}(1+M^{\prime
\prime \prime })\right) \psi _{1}+ \\
&&\kappa _{4}\left( \hbar \omega N-E+\beta -\hbar \omega \right) \left(
\gamma _{1}J_{-}+\gamma _{2}K_{+}+\gamma _{3}M^{\prime \prime \prime
}+\gamma _{4}T_{+}\right) \psi _{1}.
\end{eqnarray*}%
As we have seen the Hamiltonian can be expressed in terms of the generators
the $Sp(4,R)$ algebra without any constraints on the parameters.
Furthermore, we note that the Hamiltonians including higher order
interaction terms may also be solved within the method discussed here.

\end{document}